\def\eg{$e_g$}

\def\t2g{$t_{2g}$}
\def\lca{La$_{1-x}$Ca$_{x}$MnO$_3$}
\def\lc9{La$_{0.81}$Ca$_{0.19}$MnO$_3$}
\def\lcz{La$_{0.84}$Ca$_{0.16}$MnO$_3$}
\def\lcv{La$_{0.75}$Ca$_{0.25}$MnO$_3$}
\def\2+{$^{2+}$}
\def\3+{$^{3+}$}
\def\4+{$^{4+}$}
\def\ea{\emph{et al.}}
\def\too{$T_{O'\textrm{-}O*}$}

\documentclass[twocolumn,superscriptaddress,prl,showpacs]{revtex4}
\usepackage{graphicx}
\begin{document}

\title{Orbital order induced metal-insulator transition in \lca}

\author{Bas B. \surname{Van Aken}}
\affiliation{Solid State Chemistry Laboratory, Materials Science Centre, University of Groningen, Nijenborgh 4, 9747 AG  Groningen, the Netherlands}
\author{Auke Meetsma}
\affiliation{Solid State Chemistry Laboratory, Materials Science Centre, University of Groningen, Nijenborgh 4, 9747 AG  Groningen, the Netherlands}
\author{Y. Tomioka}
\affiliation{Correlated Electron Research Center (CERC), National Institute of Advanced Industrial Science and Technology (AIST), Tsukuba 305-8562, Japan}
\affiliation{Joint Research Centre for Atom Technology (JRCAT), National Institute of Advanced Industrial Science and Technology (AIST), Tsukuba 305-0046, Japan}
\author{Y. Tokura}
\affiliation{Correlated Electron Research Center (CERC), National Institute of Advanced Industrial Science and Technology (AIST), Tsukuba 305-8562, Japan}
\affiliation{Joint Research Centre for Atom Technology (JRCAT), National Institute of Advanced Industrial Science and Technology (AIST), Tsukuba 305-0046, Japan}
\affiliation{Department of Applied Physics, University of Tokyo,
Bunkyo-ku, Tokyo 113-8656, Japan}
\author{Thomas T. M. Palstra}
\email{palstra@chem.rug.nl}
\affiliation{Solid State Chemistry Laboratory, Materials Science Centre, University of Groningen, Nijenborgh 4, 9747 AG  Groningen, the Netherlands}

\date{\today}

\begin{abstract}
We present evidence that the insulator to metal transition in
\lca\ near $x\sim0.2$ is driven by the suppression of coherent
Jahn-Teller distortions, originating from $d$~type orbital ordering. The
orbital ordered state is characterised by large long-range $Q2$
distortions below \too. Above \too\ we find evidence for
coexistence between  an orbital-ordered and -disordered state. This
behaviour is discussed in terms of electronic phases of an orbital
ordered insulating and orbital-disordered metallic states.
\end{abstract}


\maketitle

\begin{figure}[htb]
   \centering \includegraphics[width=86mm]{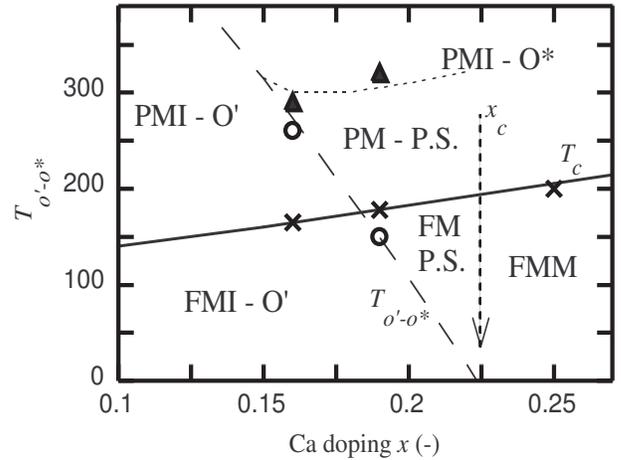} \caption{Phase diagram of \lca\ near the
   FMI-FMM transition, modified from Cheong \ea
   ~\protect\cite{Che00}. The critical concentration, $x_{\textrm{c}}$,
   indicates the metal-insulator transition at $T=0$. The phase separated region is indicated by P.S..}  \label{fig:phasediagram}
\end{figure}

LaMnO$_3$ in the ground state is an antiferromagnetic insulator with a
checkerboard pattern of \eg\
orbitals~\cite{Jon50,Hua98,Goo55,Ele71,Kug73,Cus01}. The basic
exchange interactions in the manganite perovskites allow three phases:
a ferromagnetic metal, a charge/orbital ordered antiferromagnetic
insulator and a paramagnetic polaronic liquid. Although superexchange
allows ferromagnetic interactions, the observed orbital ordering in
LaMnO$_3$ renders an overall antiferromagnetic state. When 20\% to
50\% holes are introduced, a ferromagnetic metallic ground state with
degenerate \eg\ orbitals is obtained. However, \lca, with
$0.10<x<0.20$, has a ferromagnetic insulating ground state. This
unexpected coexistence of ferromagnetic and insulating behaviour seems
to contradict the conventional double  and superexchange models. The
LaMnO$_3$-CaMnO$_3$ phase diagram by Cheong and co-workers~\cite{Che00}, sketched
partially in Fig.~\ref{fig:phasediagram}, shows the doping induced
ferromagnetic insulator (FI) to ferromagnetic metal (FM) transition at
a critical concentration of $x_c\sim0.21$. While this transition is
intriguing by itself, the situation becomes more complex by the
orbital order (O') to "not orbital ordered" (O*) transition, deduced
from anomalies in the resistivity. The origin of the coexistence of
ferromagnetism with insulating behaviour is not clear, but might stem
from a delicate balance of charge localisation by orbital ordering
(OO), due to the Jahn-Teller (JT) effect, and ferromagnetic
interactions between Mn\3+-Mn\4+. Neither the exact concentration
dependence of this transition nor the interaction of this orbital
order transition with the magnetic ordering and the temperature- or
doping-induced metal-insulator transition is known. The O'-O*
transition is typically associated with a step in the resistance or a
re-entrant insulating behaviour.

The phase diagram of Sr doped manganites has been explored in great
detail~\cite{Uru95}. Here the situation is more complicated than for Ca
doping, because the number of phases is larger due to the rhombohedral
structure at $x>0.18$ and the pronounced charge ordering (CO) at
$x\sim\frac{1}{8}$~\cite{Yam96}.  Several authors reported a JT related
structural phase transition above the magnetic ordering temperature,
$T>T_c$ at $x\sim0.12$. Below $T_c$, a transition to CO or OO is
observed, where the co-operative JT distortion is significantly reduced~\cite{Arg96,End99}. As the transition temperatures are extremely
concentration dependent, a  comparison between the various reports is
not straightforward. It is claimed that the intermediate phase is both
ferromagnetic and metallic and exhibits static co-operative JT
distortions~\cite{Kaw96b,End99}.  Some reports clearly distinguish
these two properties and combine short range order of JT distortions
with metallic behaviour~\cite{Dab99}. However, a general relation
between the JT ordered phase and the nature of the conductivity has
not been established. Also, a coincidence of the CO transition and the
re-entrant insulator-metal transition is claimed. The common
metal-insulator transition is indisputably associated with the
ferromagnetic ordering at $T_c$~\cite{Uru95,End99,Kaw96b,Dab99}.

The Ca doped phase diagram is somewhat less complex, as there is no
orthorhombic-rhombohedral structural transition. Furthermore, the
phase transitions take place at higher concentrations. As a result we
can probe the ferromagnetic insulating phase at concentrations far
away from $x=\frac{1}{8}$ to evade charge ordering. In this Letter, we explore
the region where the OO phase line crosses the magnetic ordering phase
line. We will show that the transition to the ferromagnetic metallic
phase is controlled by the suppression of JT ordering. Our
measurements show that the $Q2$ distortion is constant below
\too. However it decreases smoothly above \too, both in the
paramagnetic and in the ferromagnetic phase. We will show that the
decrease is associated with phase separation in an O' phase and an
orbital disordered (O*) phase.

\begin{figure}[htb]
   \centering \includegraphics[bb=0 610 269 800,
   width=76mm]{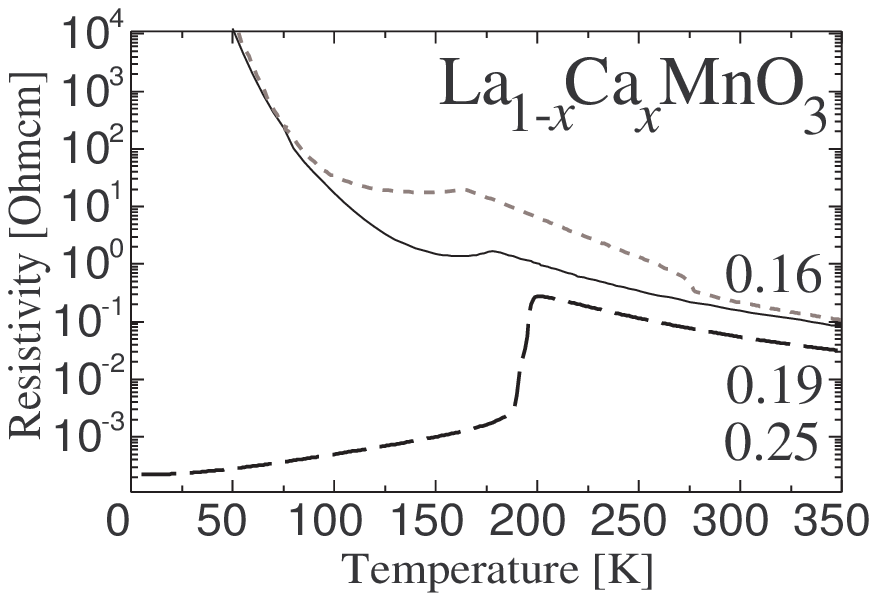} \caption{Temperature dependence of the
   resistivity of \lcz, \lc9\ and \lcv.}  \label{fig:R(T)}
\end{figure}

The experiments were carried out on single crystals of \lca, $x=0.16$,
$x=0.19$ and $x=0.25$, obtained by the floating zone method. The
sample with $x=0.19$ originated from the MISIS institute, Moscow, the
other two samples are grown at JRCAT, Japan. Although all crystals
were twinned~\cite{Van01c}, small mosaicity and sharp diffraction spots
were observed. Resistance curves for the samples were measured using a
four-point set-up. Sharp metal-insulator transitions, indicative of
the good quality of the crystals are observed for \lc9 and \lcv\ as shown in Fig.~\ref{fig:R(T)}. A
thin piece was cut from the crystals to be used for single crystal
diffractometry. Initial measurements were carried out on an
Enraf-Nonius CAD4 single crystal 4-circle diffractometer to determine
the twin relations and the twin fraction volume~\cite{Van01c}. Temperature dependent measurements between 90~K and
300~K were performed on a Bruker APEX diffractometer with an
adjustable temperature set-up.

The temperature dependence of the resistivity is shown for the three
samples in Fig.~\ref{fig:R(T)}. All three samples show a local maximum
at $T_c$. For \lcz\ and \lc9, we observe at significantly lower
temperatures, $T\approx145$~K and $T\approx160$~K respectively, that
the resistivity shows a subtle and wide transition to activated
behaviour. For \lcz\ a step in the resistance is observed at
$T\approx275$~K.

The temperature dependence of the crystal structure of \lca, $x=0.16$,
$x=0.19$ and $x=0.25$, has been determined by single crystal
diffraction. In analogy to conventional ferromagnetic metallic \lca\
systems, with $x\sim0.3$~\cite{Boo98}, we expect to observe a
narrowing of the distribution of Mn$-$O bond lengths below $T_c$ as a
result of the itinerancy in the ferromagnetic, metallic regime. As
soon as the JT orbital ordering sets in there will be an abrupt
disproportionation of the Mn$-$O bond lengths, as observed for
La$_{1-x}$Sr$_x$MnO$_3$ with $0.11<x<0.165$~\cite{Dab99}. The changes
in the structure due to orbital ordering are described using the $Q2$
distortion~\cite{Cus01}.

\begin{figure}[htb]
   \centering \includegraphics[bb=25 655 210 820]{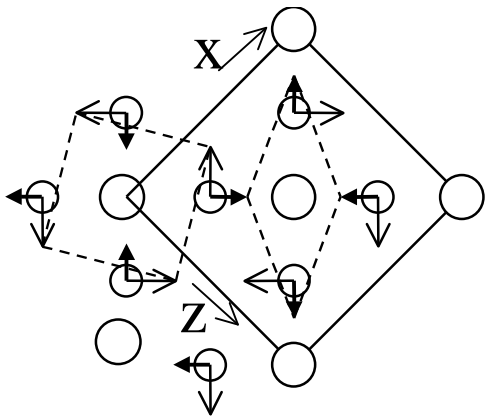}
   \caption{Sketch of the GdFeO$_3$ rotation (open arrow) and the JT
   distortion (closed arrow) in the $ac$~plane, obeying $Pnma$
   symmetry. Mn and O are represented by large and small circles,
   respectively. $Pnma$ symmetry results in a checkerboard arrangement of $Q2$ JT-distorted octahedra.}
  \label{fig:plane}
\end{figure}

The lattice parameters are not a very accurate probe to measure the
bond disproportionation, because they are the sum of long and short
bonds. However, we have shown elsewhere~\cite{Van01c} that the O2
position\cite{Q2} in $Pnma$ space group symmetry accurately reflects both the
JT distortion and the rotation of the octahedra. Because the $Q2$
distortion~\cite{Q2} and the GdFeO$_3$ rotation~\cite{Gla72} involve orthogonal
displacements of O2, they can be accurately obtained from the
fractional atomic co-ordinates of O2, as shown in
Fig.~\ref{fig:plane}. The final refinement, including the twin
relations, yielded $RF=0.068$ and $wR2=0.26$, and is  published in
detail elsewhere~\cite{VanAken}.

Fig.~\ref{fig:shift JT 3x vs t} shows the parameter for the
co-operative $Q2$ distortion against temperature as determined by
single crystal XRD. Below \too, $Q2$ is constant for the samples with
$x=0.16$ and $x=0.19$. For the $x=0.25$ sample $Q2$ is constant at all
temperatures. Because the O2 position is not constrained in $Pnma$, we
consider $Q2\sim0.002$ signalling the absence of long range JT
distortions. This value is also observed for non-Jahn-Teller active
systems such as AFeO$_3$, \emph{e.g.} $Q2=0.0036$ for LuFeO$_3$~\cite{Mar70}. We also observe in Fig.~\ref{fig:shift JT 3x vs t} that
above \too\ $Q2$ gradually decreases for $x=0.16$ and
$x=0.19$. Extrapolating the data yields that near $T\sim300$~K all
evidence for long range JT distortions is absent.

\begin{figure}[htb]
   \centering \includegraphics[bb=20 620 250 800,
   width=76mm]{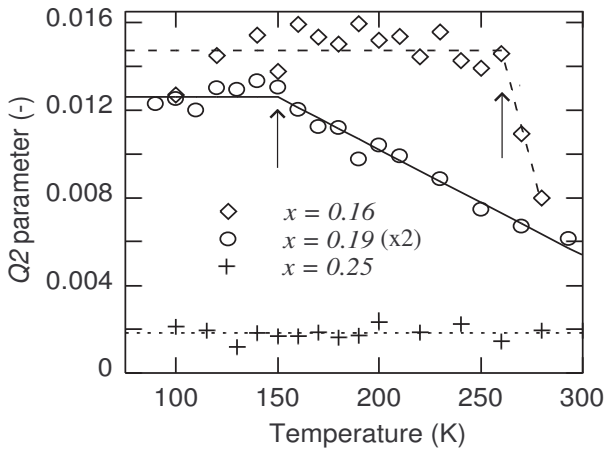} \caption{Temperature dependence of $Q2$. The data for $x=0.19$ has been multiplied by two for
   clarity. The lines are guides to the eyes. The arrows indicate 
   \too.}
  \label{fig:shift JT 3x
   vs t}
\end{figure}

Comparing the temperature dependence of the resistance with that of
$Q2$, we note that the kink in $Q2$ is accompanied with the upturn and
step in the resistance, commonly associated with \too. This is direct 
evidence that the "plateau" state in $Q2$ is the signature for the O'
phase.  These coherent distortions, associated with \eg\ orbital
ordering, are therefore sufficient to localise the charge
carriers. Thus, orbital ordering and metallicity are mutual exclusive
in the \lca\ system. This confronts the model proposed by Killian and
Khaliullin~\cite{Kil99}.

This model calculates the effect of orbital ordering on the kinetic energy of the valence electrons in terms of coherent and incoherent charge transport. An incoherent process consists of an electron that is excited to an orbital that violates the long range order. If the energy to occupy a symmetry breaking orbital is too high, the incoherent process is considered absent. The absence of incoherent processes will result in a large reduction of the holon band width, which can cause the metal-insulator transition. They argue that the reduction of the holon band width is too small to have a significant effect on the kinetic energy of the charge carriers. Thus, in their model, the orbital disorder-order crossover cannot be responsible for the metal-insulator transition. 

However, this model neglects the influence of coherent JT lattice distortions. An incoherent process not only involves the orbital excitation energy associated with the energy difference of the two \eg\ orbitals, but also compromises the $Pnma$ symmetry that incorporates the coherent $Q2$ distortions. Therefore, incoherent processes do not only involve crystal field energies, but require adjustment of the local oxygen co-ordination such that the orbitals and lattice distortions are aligned. Our main finding is that orbital ordering has a large effect on electronic conduction and moreover that metallicity only results in \lca\ if the orbital degeneracy is maintained.

When holes are introduced on JT distorted Mn\3+ sites, the resulting
Mn\4+ ions still experience a JT distorted oxygen co-ordination. The
reason is that the perovskite lattice consists of corner sharing
oxygen octahedra. Therefore, the oxygen position is determined by two Mn ions. Thus a Mn\4+ ion co-ordinated by Mn\3+ containing octahedra will still have a distortion, albeit with a smaller amplitude. This is in contrast to single ions models~\cite{Kil99}, that neglect the co-operative effect of the ordered $Q2$ distortions.

\begin{figure}[htb]
   \centering 
\includegraphics[bb=20 614 213 800, width=76mm]{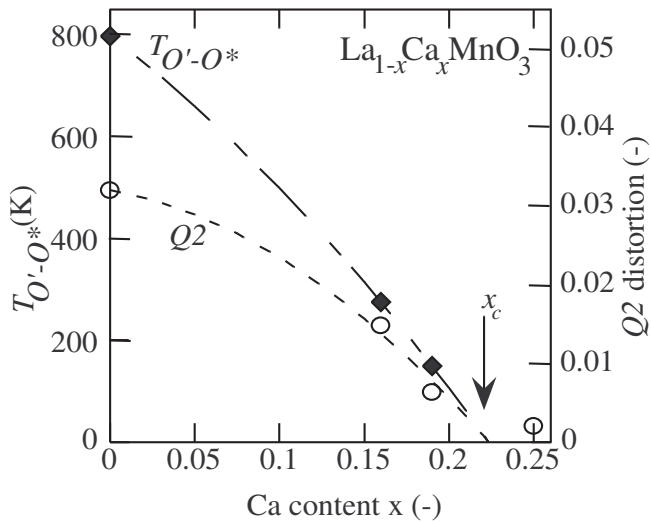} 
\caption{\too\ and $Q2$ against Ca concentration. The $Q2$ value for LaMnO$_3$ ($x=0$) has been taken at 300~K~\cite{Rod98}, \too\ is taken from  Ref.\protect\cite{Mur98b}. The dashed lines are a guide to the eye.}
  \label{fig:q2 vs x}
\end{figure}

Nevertheless, introduction of holes in orbital ordered LaMnO$_3$ will decrease the magnitude of the $Q2$ distortion. Our experimental values at 100 K are shown in Fig.~\ref{fig:q2 vs x}. Clearly $Q2$ decreases gradually with Ca doping and no JT distortion can be observed for $x>0.21$. The $Q2$ parameter for the undoped \eg\ system LaMnO$_3$ is about four times larger than in undoped \t2g systems such as YVO$_3$~\cite{Bla01} and YTiO$_3$~\cite{Mac79}. Furthermore, \too\ is reduced with increased doping level. The reduction in $Q2$ and \too\ with increased doping level is consistent with resistivity and Seebeck measurements, which showed that the activation energies for charge transport decreased with doping~\cite{Pal97}. We speculate that the disappearance of long-range orbital order is partially induced by frustrating the $d$~type orbital ordering by introducing holes with doping. Undoped LaMnO$_3$ has antiferromagnetic interactions along the $b$ axis, which is consistent with mirror symmetry perpendicular to the $b$~axis and the ferrodistortive orientation of the orbitals along the $b$~axis. Introduction of holes will result in ferromagnetic interactions along $b$, resulting eventually for $x>0.10$ in a ferromagnetic ground state. However, a larger carrier concentration is required to suppress the orbital order and obtain degeneracy of the \eg\ orbitals and thus a metallic state.

Whereas conventionally \lca\ is considered to be a doped antiferromagnet in which double exchange plays a dominant role, we emphasise the dominating role of introducing holes in the orbital ordered state.

We note that the temperature dependence of the $Q2$ distortion is remarkably different from that observed in \t2g based JT ordered systems. For these materials the JT distortion exhibits a BCS-like type temperature dependence, with the vanishing of the coherent distortion above \too. Here, we observe a rapid decrease of the coherent distortion above \too~\cite{Bla01}. We interpret this temperature dependence originating from a coexistence of an orbital ordered and orbital disordered state. We note that the measurements of integrated intensities cannot give more detail of the nature of these states. This coexistence was also observed in neutron powder diffraction experiments on La$_{0.86}$Ca$_{0.16}$MnO$_3$ at room temperature~\cite{Dab99b}. Such structural phase separation is evidence for electronic phase separation as we associate the orbital ordered state with localised charge carriers and the orbital degenerate state with, if $T<T_c$, the metallic state.

In the phase diagram of La$_{1-x}$Sr$_{x}$MnO$_3$, the CO phase
borders the FMM phase, as observed by superlattice reflections in
single crystal neutron experiments~\cite{Yam96}. In contrast, for
\lca\ the CO phase is suppressed by the orbital ordered FMI phase. We
have not observed any superlattice reflections. A possible charge
ordering phase either exists at lower temperatures, $T<90$~K, or at a
hole concentration closer to $x=1/8$. We note that the concept of
orbital polarons might lead to low temperature charge and orbital
ordering~\cite{Kil99,Miz01}. However this feature is incompatible with
$Pnma$ symmetry, and thus can be ruled out.

We have demonstrated that the ferromagnetic metallic phase is obtained
by the suppression of the long range Jahn-Teller ordering. This
contrasts the common opinion that metallicity occurs if the
charge carrier density exceeds a critical concentration. The O' phase
mixes with the O* phase above \too. Above a second phase line, the
transition to the O* phase is complete. The metallic state of \lca\ is
bounded by ferromagnetic ordering and the absence of orbital ordering.

Stimulating discussions with Daniel Khomskii, Lou-F\'{e} Feiner, George
Sawatzky, Graeme Blake, Martine Hennion, Paolo Radaelli, Neil Mathur and Takashi Mizokawa are gratefully acknowledged. This work is supported by the Netherlands Foundation for the Fundamental Research on Matter (FOM) and by the New Energy and Industrial Technology Development Organization (NEDO) of Japan.


\begin{thebibliography}{27}
\bibitem[{Q2()}]{Q2}
\bibinfo{note}{$Q2$ is defined as $x_{O2}+z_{O2}-\frac{1}{2}$. The O2 position,
  ($x_{O2}, y_{O2}, z_{O2}$), near ($\frac{1}{4}, 0, \frac{1}{4}$) is used
  here.}
\end{thebibliography}
\end{document}